\documentclass[useAMS,usenatbib]{mn2e}
\usepackage{latexsym,graphicx,natbib,psfig}

\newcommand{\be} {\begin{equation}}

\newcommand{\ee} {\end{equation}}

\newcommand{\src}{SAX\,J1808.4-3658}

\newcommand{\XMM}{{\em XMM--Newton}}

\newcommand{\bc}{\begin{center}}
\newcommand{\ec}{\end{center}}
\def\ltsima{$\; \buildrel < \over \sim \;$}
\def\lsim{\lower.5ex\hbox{\ltsima}} 
\def\loe{\lower.5ex\hbox{\ltsima}}
\def\gtsima{$\; \buildrel > \over \sim \;$}
\def\gsim{\lower.5ex\hbox{\gtsima}}
\def\goe{\lower.5ex\hbox{\gtsima}}

\def\ergscm2 {erg\,s$^{-1}$cm$^{-2}$}

\title[XMM-Newton observations of \src]{\src\,: high resolution spectroscopy and decrease of pulsed fraction at low energies}

\author[Patruno et al.]{Alessandro Patruno$^{1}$\thanks{email: a.patruno@uva.nl},
 Nanda Rea$^{1}$, Diego Altamirano$^{1}$, Manuel Linares$^{1}$,\and
R. Wijnands$^{1}$ and M. van der Klis$^{1}$\\
$^{1}$Astronomical Institute ``A. Pannekoek'', University of Amsterdam, Kruislaan 403, 1098 SJ, The Netherlands.\\
}

\begin{document}

\maketitle

\begin{abstract}

{{\it XMM-Newton}} observed the accreting millisecond pulsar SAX
J1808.4-3658 during its 2008 outburst. We present timing and spectral
analyses of this observation, in particular the first pulse profile
study below 2 keV, and the high-resolution spectral analysis of this
source during the outburst. Combined spectral and pulse profile
analyses suggest the presence of a strong unpulsed source below 2 keV
that strongly reduces the pulsed fraction and a hard pulsed component
that generates markedly double peaked profiles at higher energies.  We
also studied the high-resolution grating spectrum of SAX J1808.4-3658,
and found several absorption edges and Oxygen absorption lines with
whom we infer, in a model independent way, the interstellar column
densities of several elements in the direction of SAX J1808.4-3658.

\end{abstract}


\section{Introduction}

The accreting millisecond X-ray pulsar (AMXP) SAX J1808.4--3658 (J1808
from now on) was the first X-ray binary found to pulsate in the
millisecond range, with a spin period of 2.5 ms \citep{wij98}). It has
been observed in outburst 6 times, roughly every 2.5 years since 1996.
During the outburst, the magnetic field is thought to channel part of
the disc material onto the neutron star magnetic poles.  The radiation
emitted from the impact region (hot spot) and/or a slab of shocked
material above it is then modulated at the neutron star spin
period. This radiation is observed as pulsed emission that adds to the
unpulsed emission coming from the accretion disc. A possible
comptonizing medium surrounding the impact region can upscatter part
of the radiation to higher energies \citep{pou03}.  The pulsations and
the X-ray spectrum were observed during previous outbursts by {{\it
RXTE}} and a first study of the 1998 outburst was performed using
those data (\citealt{pou03}, \citealt{ibr09}). J1808 was never
observed below 2 keV during an outburst (except in the 2000 and 2005
outbursts at very low luminosity levels, see \citealt{wij03} and
\citealt{cam08}).  This energy range is important to understand the
pulse formation mechanism, because it is here that both the hot spot
and the accretion disc thermal emissions are expected to peak.  Also,
many absorption lines from the interstellar medium are expected in
this energy range. These lines are important to definitively determine
the interstellar column density toward the source.  Broadband spectral
analyses of this {{\it XMM}} observation were reported by
\citet{pap09} and \citet{cac09}, who both focused on the study of the
iron line emission at 6--7 keV.  Here we present the first
simultaneous spectral and timing analysis of the pulsations of J1808
as observed with {{\it XMM-Newton} during the 2008 outburst, with a
particular attention to the lower energy range ($<2$ keV).

%%%%%%%%%%%%%%%%%%%%% FIGURE FA%%%%%%%%%%%%%%%%%%%%%%%%%
\begin{figure*}
  \begin{center}
\hbox{
    \rotatebox{-90}{\includegraphics[width=0.7\columnwidth]{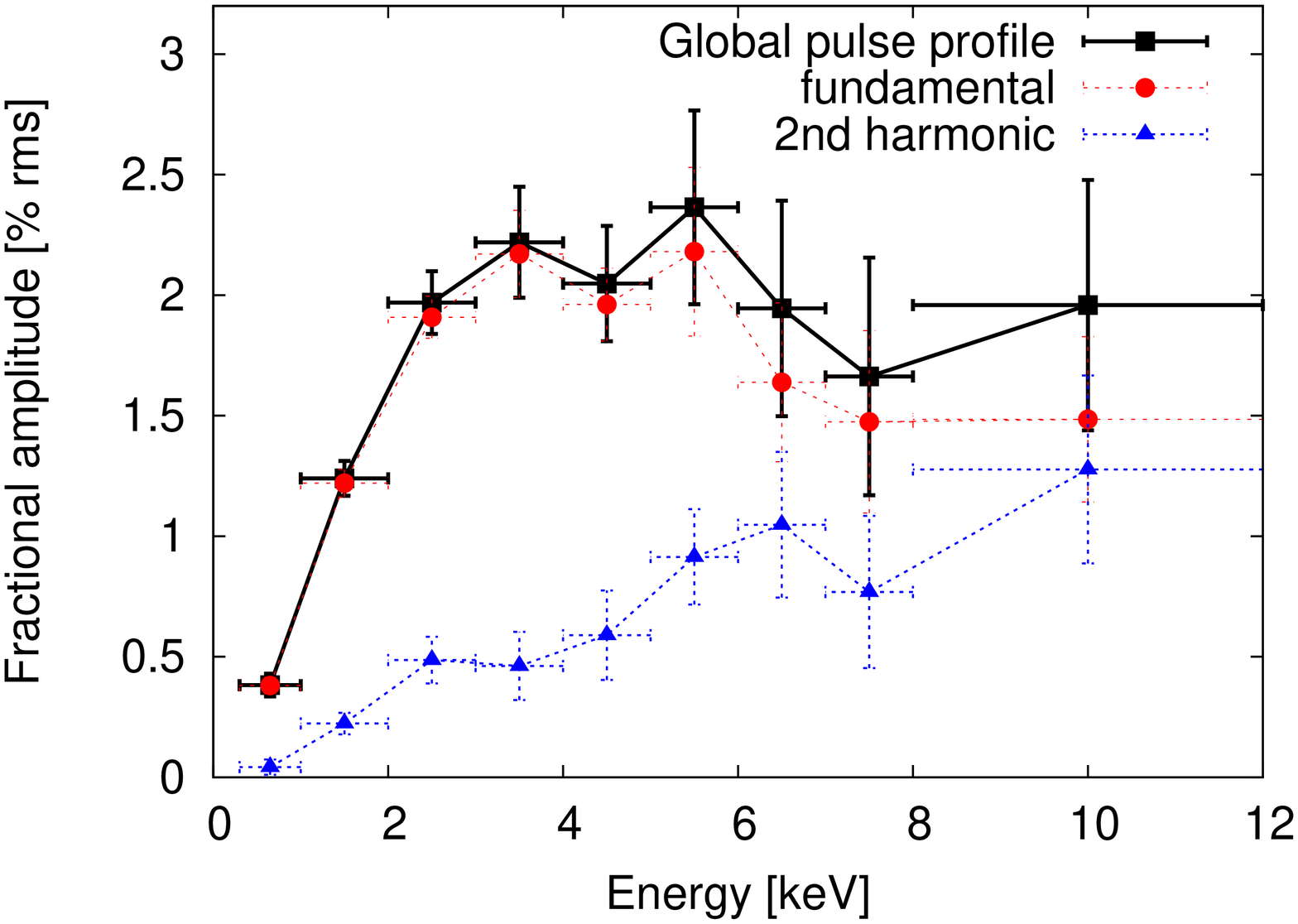}}
    \rotatebox{-90}{\includegraphics[width=0.7\columnwidth]{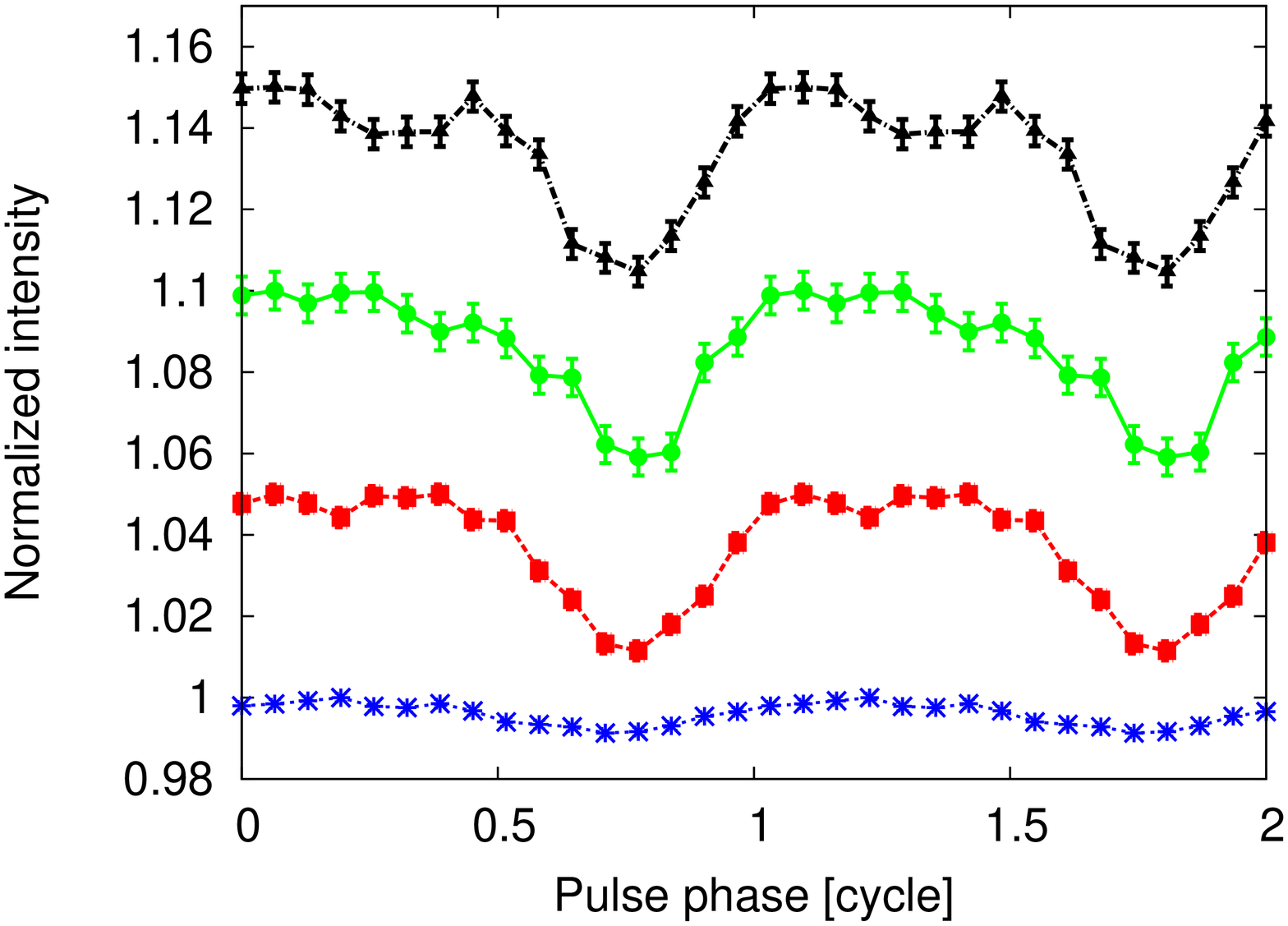}}}
%    \rotatebox{-90}{\includegraphics[width=0.5\columnwidth]{flux-res.eps}}}
  \end{center}
  \caption{\textbf{Left panel}: fractional amplitude for the pulse
profiles in 9 energy bands.  The circles and triangles refer to the
fundamental and second harmonic respectively. Vertical bars indicate
$1~\sigma$ errors in amplitude and horizontal bars the energy range.
The black squares are the fractional amplitudes of the global profile,
with the rms of the fundamental and second harmonic added in
quadrature.  The pulsed fraction is much lower below $\approx 2$ keV,
reaching a minimum value of $\approx 0.4\%$ rms in the 0.3-1 keV
energy range.  \textbf{Right panel}: pulse profiles in four different
energy bands (blue asterisks 0.3--2 keV, red squares: 2--3.5 keV,
green circles: 3.5--4.5 keV, black triangles 4.5--12 keV). The
profiles are double peaked, with the strength of the second harmonic
increasing at higher energies.  Two cycles are plotted, with profiles
normalized to the maximum intensity and shifted by arbitrary
amounts. The bottom profile corresponds to the low energies (0.3-2
keV) and its fractional amplitude is considerably reduced (although
still significantly detected).}
\label{fig:profiles} 
\label{fig:flux-res}
\label{fig:framp}
\end{figure*}

%%%%%%%%%%%%%%%%%%%%%%%%%%%%%%%%%%%%%%%%%%%%%%%%%%%%%%%%%%%%%%%%

\section{X-ray observation}
\label{obs}

J1808 was observed in outburst with \XMM\ on 2008 October 1st
(MJD 54740), for 63 ks of on--source exposure time.  At the time of
this observation, J1808 was at the beginning of the exponential decay
stage of the outburst, with a relatively high flux level (see Hartman
et al. 2009 for a description of the overall outburst lightcurve).
The \XMM\, Observatory \citep{jan01} includes three 1500~cm$^2$ X-ray
telescopes with the European Photon Imaging Camera (EPIC), and a
Reflecting Grating Spectrometer (RGS; \citealt{den01}). An Optical
Monitor is also present (\citealt{mas01}). It is used to follow
optical counterparts and will not be considered in this work.  The
EPIC camera is composed of two MOS CCDs \citep{tur01} and a pn CCD
\citep{str01}.  Each EPIC camera has a fixed, mode dependent frame
read-out frequency, producing event lists in the 0.1-12 keV energy
range. The RGS, is composed of a double array of gratings and produces
high resolution spectra in the 0.33 to 2.5 keV energy range.

Data have been processed using SAS version 8.0.1, and we have employed
the most recent calibration files (CCF) available at the time the
reduction was performed (February 2009). Standard data screening
criteria were applied in the extraction of scientific products. After
removing solar flares and telemetry dropouts the net pn exposure time
is 41\,ks. 
The central CCD of MOS1 was operated in {\tt full frame} mode with
{\tt thin} filters, and is heavily piled--up. For this reason we do
not consider the MOS1 data any further.  The MOS2 was operated in {\tt
timing window} mode, and is also excluded from our analysis since the
1.5\,ms time resolution is insufficient to study the pulsations, and
its spectral capabilities for very bright sources are not as much
calibrated as the pn camera. The pn camera was operating in {\tt
timing} mode (with a {\tt thin} filter),in order to reduce pile--up
and allow the high precision timing analysis required for an accreting
millisecond X-ray pulsar (AMXP).
We extracted the source photons from the pn with RAWX coordinates
26-49. The background is obtained from a region of the
same size, at RAWX 2-25. Only photons with PATTERN$\leq 4$ were used.
The extracted spectrum was rebinned before fitting to obtain at
least 100 counts per bin and the pn energy resolution was not oversampled
by more than a factor three. We also extracted first and second order
RGS1 and RGS2 spectra, using the standard procedure reported in the
\XMM\, analysis manual, and again we ensured to have at least 100
counts per spectral bin in any RGS spectrum.

\section{Timing analysis}
\label{timing}

We have first corrected the event times to the barycenter of the Solar
System (using the SAS tool {\tt barycen}, and the optical position
given in \citealt{har08}) and then we applied the 2008 outburst timing
solution published in \citet{har09} in order to predict the phases of
each photon detected and reconstruct the pulse profiles (see
\citealt{pat09} for a detailed explanation of the timing technique).
For the timing analysis we use events in the 0.3--12\,keV energy
range.  The pulse profiles are built by folding data segments of
length $\sim 3500$ s for the pulse phase analysis. This length is
chosen to guarantee sufficiently high signal-to-noise profiles even if
the pulsed fractions are small. The pulses are then decomposed by
fitting two harmonics with frequency fixed at the pulse frequency
(fundamental, $\nu$) and twice the pulse frequency (second
harmonic, $2\nu$) plus a constant representing the non-pulsed emission.

We did not attempt to calculate a new timing solution since the
precision of the solution achievable with the short observation
baseline of {{\it XMM}} is at least an order of magnitude lower than
what was obtained with the {{\it RXTE}} data \citep{har09}.  The short
baseline of the observation is also insufficient to model the timing
noise that affects the pulse phases and that was extensively discussed
in \citet{har08, har09} for J1808.  If timing noise is present, a
systematic error is introduced in the determination of the pulse
phases and spin frequency \citep{pat09}.

Therefore we decided to subtract the solution reported in
\citet{har09} and obtain the phase residuals with respect to that
constant pulse frequency plus Keplerian circular orbit model.  The
pulse phase residuals of the fundamental drift by $\approx 0.1$ cycles
during the observation. We also found a correlation between these
pulse phase residuals and the 0.3-12 keV X-ray flux.  We fitted the
data with a linear relation that gives a $\chi^{2}$ of 11.4 for 10
degrees of freedom, and a slope of $(9.9\pm1.2)\times10^{-4}\rm\,
cycle/ct/s$ (we quote the 1$\sigma$ error).

The second harmonic is significantly detected in only $\approx 1/3$ of
the profiles, where a detection is defined as a ratio between the
pulse amplitude and its statistical error larger than 3.  When not
detected, the second harmonic fractional amplitude upper limit was
between 0.4 and 0.6$\%$ rms at the 98$\%$ confidence level.  No
correlation between pulse phases and flux was found for the second
harmonic.

To increase the signal to noise and calculate the harmonic content of
the pulsations in the whole energy band, we folded the entire 41 ks
data into one single pulse profile.  The fractional rms amplitude of
fundamental and second harmonic in the 0.3-12 keV energy band is
0.98(2)\% rms and 0.33(3)\% rms respectively (1$\sigma$
uncertainties).

We then repeated the procedure by dividing the observations in 9
energy bands between 0.3 and 12 keV.  The fractional amplitude of the
pulse profile increases from 0.3 up to 3 keV, and then it remains
constant within the errors up to 12 keV (Fig~\ref{fig:framp}).  The
fundamental tracks the behaviour of the total pulse profile.  The
second harmonic increases monotonically in the energy range
considered.  At energies above $\approx 6$keV the fractional
amplitudes of the fundamental and second harmonic are comparable and
the overall pulse profile is double peaked (Fig.~\ref{fig:profiles},
see also \citealt{har09}).

\section{Spectral analysis}
\label{spectral}

We performed spectral analysis using the EPIC-pn spectrum (extracted
as reported in \S~2) in the 0.6--12\,keV energy range, and the RGS1
and RGS2 in the 0.4--1.8\,keV energy range for the first order, and
0.7--1.8\,keV for the second order. {\tt XSPEC} version 11.3 was used
for the spectral analysis. 

To account for relative flux calibration uncertainties between different
instruments, we fitted a multiplicative constant with the model,
allowing for up to 10\% relative calibration flux scaling between
EPIC-pn and RGS\footnote{http://xmm2.esac.esa.int/docs/documents/CAL-TN-0052-5-0.ps.gz}.
The relative offsets between the instruments are found to be less than
4\% in all our fits. Furthermore, we included a 1.5\% systematic to
the errors of each spectral bin (using the relative {\tt XSPEC} tool)
to take into account the calibration inaccuracies of each single
instrument used\footnote{http://xmm2.esac.esa.int/external/xmm\_sw\_cal/calib}.
We first used solar abundances from \citet{and89} and cross-sections
from \citet{bal92} for the photoelectric absorption.  We tried an
absorbed power-law plus a multi temperature disc and a single
temperature blackbody model ({\tt phabs$*$(diskbb + bbody +
powerlaw})) as suggested in \citet{cac09} and \citet{pap09}.  The
$\chi_{\nu}^2 = 5$ was unacceptable, mainly because of unmodelled
features in the data points between 0.3--2\,keV and 6--7\,keV.

The 6-7\,keV energy range is where fluorescence lines of Fe are
expected \citep{geo91}. Following \citet{cac09} and \citet{pap09} who
claimed the detection of a broad iron $\rm\,K_{\alpha}$ line in this
energy range, we fitted this feature with a {\tt diskline} model (see
Tab.\,1).  We refer to \citet{cac09} and \citet{pap09} for discussion
of this broad iron line.  To model the 0.3--2\,keV features, we first
tried several photoelectric cross-sections, and different element
abundances.  The residuals are not very sensitive to the photoelectric
cross-section parameters, while they strongly depend on the assumed
abundances. The best fit model, however not yet satisfactorily, was
found using the \citet{bal92} cross-sections, and the \citet{wil00}
abundances.  The stronger low energy features in the residuals were
coming from absorption and emission features close to the oxygen
K-edge ($O_{K}$-edge) at 0.543\,keV.  We decided to model only the
continuum and the iron $\rm\,K_{\alpha}$ line as a first step,
ignoring the data between 0.5-0.6\,keV. This energy range and the
single features it contains were then investigated separately by using
the RGS data (see \S~\ref{lines}, and Table\,1 for the continuum and
Fe line spectral results).

The two weak features at 1.8\,keV and 2.2\,keV
are known to be due to the instrumental Si and Au K edges, not yet
perfectly calibrated, especially when dealing with {\tt timing} mode
observations (see the EPIC calibration report in the footnote).  We do
not find evidence of a 0.871\,keV O VII edge (reported in
\citealt{pap09}), and no other significant edges are detected in the
0.8-1 keV range.

After removing the O-Si-Au edges (0.5-0.6 and 1.6-2.3\,keV), we obtain
$\chi_{\nu}^2 =1.41$ (4051 dof).  Given the high quality X-ray
spectrum, the relatively high value of $\chi_{\nu}^2 =1.41$ is very
likely due to inter-calibration problems between the pn and the RGS
spectra, and to the calibration uncertainties of each camera which
emerge when observing bright sources.  In fact, when using only the pn
spectrum, we obtain a statistically acceptable fit with a
$\chi_{\nu}^2 =0.95$ for 235 dof, with the same spectral parameters as
reported in Tab.\,1. The same model applied to the RGS alone gives a
$\chi_{\nu}^2 =1.1$. Therefore we accept the $\chi_{\nu}^2 =1.41$ and do not
further complicate the spectral model.

\begin{figure}
  \begin{center}
    \hbox{
%      \rotatebox{-90}{\includegraphics[width=0.45\textwidth]{res_spectrum_total_noedges_new.ps}}
      \rotatebox{0}{\includegraphics[width=0.90\columnwidth]{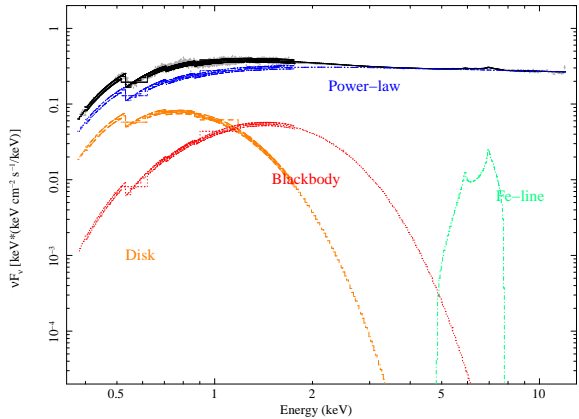}}}     
  \end{center}
  \caption{%\textbf{%Left panel}: from top to bottom: EPIC-pn and RGS1 and RGS2
%  first and second order spectra for \src; residuals of the best fit
%  without modeling the Si-K and Au-K instrumental lines and the oxygen
%  forest; residuals obtained when cutting away all the unmodelled lines
%  (continuum spectral parameters do not change).~\textbf{Right
%  panel}:
  $\nu$F$_{\nu}$ plot of the model used to fit the continuum. 
  The power-law flux dominates over the other spectral components.}
  \label{fig:vfv}
\end{figure}

\subsection{High-resolution spectroscopy}
\label{lines}

We  inferred  the interstellar  medium  (ISM)  abundances  in a  model
independent way, by separately fitting the absorption edges to the RGS
data. We  used only the RGS1  and RGS2 first order  spectra, which are
the best calibrated and have  a higher number of counts. The continuum
parameters were kept fixed at the value reported  in Table\,1. We used
{\tt  vphabs}  that allows  to  set  fixed  abundance parameters  with
respect to the solar composition  and to isolate the single absorption
edges. The  strongest features were observed around  the $O_{K}$-edge
($O_K$; see Fig.\ref{figlines}).  To model  this edge we fixed at zero
the oxygen abundance of the {\tt  vphabs} model, and fit only the data
around the $O_K$ with an {\tt  edge} model (note that the continuum is
relatively constant in the small  energy range around the edge itself,
hence  the  modelling  is   independent  on  the  broadband  continuum
model).The best fit gives $O_K = 0.5421\pm0.0003\rm\, keV$ and $\tau =
0.66\pm0.01$,   with  $\tau_x=N_{x}\times\sigma$   and   $\sigma$  the
photoelectric cross section. With the same method we fitted also the
iron L ($Fe_{L}$), neon  K ($Ne_K$), magnesium K ($Mg_K$)  and silicon K ($Si_K$)
edges that lie in the RGS band. The column density for each element is
reported in Tab.\ref{tablines}.
The absorption features close to $O_K$ were fitted with Gaussian lines
and were identified as $\rm 1s-2p$ atomic transitions of O\,I, O\,II
and O\,III.
%at $0.527\pm0.001$\,keV (with an equivalent width of
%$EQW=1.2\pm0.2$\,eV), O\,II at $0.531\pm0.001$\,keV
%($EQW=1.2\pm0.2$\,eV), and O\,III at $0.537\pm0.001$\,keV
%($EQW=0.6\pm0.1$\,eV). 
The O\,IV line lies too close to an instrumental bad
column to be detectable. The levels higher than IV are not detected,
with a 98\% upper limit of $0.08$ eV on the equivalent width (EW).
Similar features have been observed in other X-ray binaries like
XB\,1254--690 \citep{dia09} and Cyg X-2 (\citealt{tak02};
\citealt{cos05}).

From the single absorption edges we calculated the column density of
each element ($\tau_x=N_{x}\times\sigma$) assuming photoelectric cross
sections from \citet{gou91}. The equivalent hydrogen column density is
inferred from the best measured edge ($O_{K}$), by using abundances
from \citet{wil00}. We derived $N_H
=(1.4\pm0.2)\times10^{21}$\,cm$^{-2}$, consistent with the value
derived in the direction of J1808 from both HI and HII measurements
($1.3\times10^{21}$\,cm$^{-2}$ and $1.14\times10^{21}$\,cm$^{-2}$;
\citealt{dic90} and \citealt{kal05} respectively).

From the measurement of the EW of the oxygen I, II, and III $\rm
1s-2p$ transitions, using the method of the curve of growth, we have
an independent measure of the relative oxygen abundances, using the
formula $EW_{\lambda} =8.85\times10^{-13} N_{x}\lambda^2 f_{ij}$,
where $\lambda$ is the wavelength of the line, and $f_{ij}$ the
oscillator strength for the transition \citep{spi78}. We found an
$N_{\rm oxygen}$ consistent with that inferred from the $O_{K}$-edge
(see Tab.\,2).

%%%%%%%%%%%%%%%%%%%%% FIGURE vFv  %%%%%%%%%%%%%%%%%%%%%%%%%
\begin{figure}
  \begin{center}
\vbox{
    \rotatebox{-90}{\includegraphics[width=0.7\columnwidth]{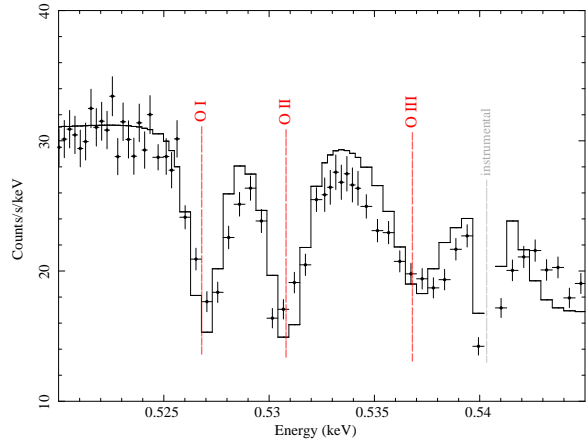}}}
  \end{center}
  \caption{High resolution RGS1 spectrum of J1808 around the $O_{K}$-edge:
OI, OII and OIII absorption lines for the 1s-2p transition.
The O IV line falls too close to the instrumental edge to be measured.}
    \label{figlines}
\end{figure}

%%%%%%%%%%%%%%%%%%%%%%%%%%%%%%%%%%%%%%%%%%%%%%%%%%%%%%%%%%%%%%%%

%%%%%%%%%%%%%%%%%%%%%%%%% Spectral parameters CONTINUUM%%%%%%%%%%%%%%%%%%%%%%%%%%%%%%%%%%%%%
\begin{table}
\begin{tabular}{lcc}
\hline
\hline
\multicolumn{1}{l}{Parameters} & DiskBB+BB+PL+Diskline \\
\hline
N$_{H}$ ($10^{22}$\,cm$^{-2}$) & $0.16\pm0.02$  \\
Inner Disc kT (keV) &  $0.20\pm0.01$ \\
Disc flux  (\ergscm2 ) & $(1.89\pm0.21)\times10^{-10}$ \\
$\rm\,kT_{BB}$ (keV) &  $0.33\pm0.01$  \\
BB radius (km) & $10.6\pm3.2$ \\
BB flux  (\ergscm2 ) & $(1.22\pm0.15)\times10^{-10}$ \\
Photon Index $\Gamma$ & $2.11\pm0.01$ \\
PL flux   (\ergscm2 ) & $(1.53\pm0.09)\times10^{-9}$ \\
E$_Fe$ (keV) & $6.45\pm0.08$  \\
EW (eV) & $97.7\pm31.4$ \\
R$_{IN}$  (km) & $20\pm2$  \\
R$_{OUT}$ (km) &  $193\pm15$ \\
Incl. (deg) & $>44^{a}$  \\
$\beta^{b}$ & $-2.1\pm0.1$ \\  
Fe-line flux  (\ergscm2 ) & $(6.8\pm0.2)\times10^{-12}$ \\
\hline
Flux  (\ergscm2 ) & $(1.85\pm0.08)\times10^{-9}$ \\
Absorbed Flux (\ergscm2 ) & $(1.51\pm0.08)\times10^{-9}$\\
%L$_{\rm bol}$  (\ergs ) & $(1.7\pm0.2)\times10^{37}$ \\
\hline
$\chi^2_{\nu}$ (dof) & 1.41 (4051)  \\
\hline
\hline
\end{tabular}
\caption{Spectral parameters for J1808 in outburst, from combined pn and all RGS data. 
Errors are at 1$\sigma$ confidence level. N$_{H}$ is calculated with abundances from \citet{wil00}. 
Unabsorbed fluxes are given in the 0.5-10\,keV energy range.
The blackbody radius is calculated assuming a distance of 3.5\,kpc and
a neutron star mass of $1.4M_{\odot}$.
\textbf{a.} The lower limit is quoted at 95$\%$ confidence level.
\textbf{b.} $\beta$ is the power law index of the emissivity}
\label{spec}
\end{table}

%%%%%%%%%%%%%%%%%%%%%%%%%%%%%%%%%%%%%%%%%%%%%%%%%%%%%%%%%%%%%%%%%%%%%%%%%%%%%

%%%%%%%%%%%%%%%%%%%%%%%%% Spectral parameters RGS %%%%%%%%%%%%%%%%%%%%%%%%%%%
\begin{table}
\begin{tabular}{lcccc}
\hline
\hline
Edge & Energy (keV) & $\tau$ & $\sigma$ & $N_{x}$ \\
\hline 
$O_K$ &  $0.5421\pm0.0003$ & $0.66\pm0.01$ & 5.642 & $11.7\pm0.2$ \\
$Fe_L$ & $0.712\pm0.005$ & $0.08\pm0.02$ & 4.936 & $1.6\pm0.4$ \\
$Ne_K$ &  $0.865\pm0.004$ & $0.10\pm0.02$ & 3.523 & $2.8\pm0.5$ \\
$Mg_K$ & $1.281\pm0.007$ & $0.06\pm0.01$ & 2.191 & $2.7\pm0.4$ \\
$Si_K$ & $1.79\pm0.007$ & $0.14\pm0.01$ & 1.476 & $9.4\pm0.6$ \\
\hline
\hline
1s-2p & Energy (keV) & EW (eV)& $N_{oxygen}$& \\
\hline 
$O I$ &  $0.527\pm0.001$ & $1.2\pm0.2$ & $9\pm2$ \\
$O II$ &  $0.531\pm0.001$ & $1.2\pm0.2$ & $8\pm2$ \\
$O III$ &  $0.537\pm0.001$ & $0.6\pm0.1$ & $11\pm2$ \\
\hline
\end{tabular}
\caption{Interstellar medium edges and lines in the direction of \src. Photoelectric cross sections ($\sigma$) are in units of
$10^{-19}$\,cm$^{2}$ (from Gould \& Jung 1991), while the column densities $N_{x}$ are in units of $10^{17}$cm$^{-2}$.}  
\label{tablines} 
\end{table}
%%%%%%%%%%%%%%%%%%%%%%%%%%%%%%%%%%%%%%%%%%%%%%%%%%%%%%%%%%%%%%%%%%%%%%%%%%%%%%

 \section{Discussion} \label{discussion}

We observed for the first time the pulsations in the 0.3-2 keV energy
range for J1808.  The pulse fractional amplitudes sharply decrease
below $\approx 2$keV during the 2008 \XMM\, observation. Our spectral
analysis required a multi temperature blackbody at low energies that
we identify with the accretion disc emission, as well as the main
source of unpulsed emission. 
We interpret the higher energy
blackbody as the emission coming from the hot spot of the neutron
star.  The power law is also a pulsed component, since we measured
$\approx 2\%$ rms pulse amplitudes up to 12 keV were the power law
emission dominates.

  The empirical power-law model requires a disc contributing only
$\approx 30\%$ of the power-law flux below 2 keV (Fig~\ref{fig:vfv}).
Therefore the disc is not the only responsible of the steep decrease
of pulse fractional amplitudes below 2 keV. However, an empirical
power-law model is unphysical and a self consistent physical scenario
would require a sharp decrease of the power-law component below
$\approx 2-3$ keV.  In that case the unpulsed disc emission might
dominate below 2 keV and explain the sudden drop of the pulse
amplitudes at those energies.  A Comptonization model with a shocked
plasma in a slab geometry is expected to cut off below $\approx3$ keV
and was already proposed by \citet{gie05} for the AMXP XTE J1751-305.
The hot single temperature blackbody would then be produced by the
pulsating emission of the hot spot, the multi temperature blackbody by
the unpulsed radiation of the accretion disc and the hard component by
the pulsed comptonized radiation of the shock around the hot spot.

The origin of the second harmonic can be related to a different pulse
profile of the hard emission compared to the lower-energy blackbody as
was suggested by \citet{gie02} and \citet{pou03} for {{\it RXTE}}
observations of J1808, and by \citet{gie05} for {{\it XMM}} and {\it
RXTE}} observations of XTE J1751-305.  This might result from a
different angular distribution of the Comptonized radiation, which is
expected as it is produced in the optically thin accretion shock, but
not at the stellar surface as the blackbody emission.

The reason why the phase of the fundamental is correlated with the
X-ray flux while the second harmonic is not, can then be related with
different formation processes for the fundamental and the second
harmonic.  If the hot spot contributes only to the fundamental
frequency while the comptonization region contributes to both the
fundamental and the second harmonic, the fundamental pulse phase may
track the hot spot position.  The phase of the second harmonic instead
will be affected by the comptonization process and might come from an
extended region around the hot spot, covering a large area of the
neutron star surface.
The hot spot can instead come from a well defined region on the
neutron star surface, and can move according to the X-ray flux
fluctuations thus producing the pulse phase wandering correlated with
X-ray flux (\citealt{rom04},~\citealt{lam08}).

We also measured the first model independent column densities of
several elements in the line of sight of J1808 (Tab.\,2). The most
precise measurement comes from the oxygen column density, from which
we could derive the equivalent hydrogen column density
$N_{H}=1.4\times10^{21}$\,cm$^{-2}$, assuming abundances from
\citealt{wil00}. This determination is particularly important: i) in
the study of the X-ray emission of this object during quiescence and
cooling, where the uncertainty in the assumed $N_{H}$ value could
alter the reliability of the source intrinsic luminosity (see for
example \citealt{hei09} for cooling studies of \src\, and
\citealt{yak04} for a discussion of the problem), and ii) for optical
studies who can now rely on a more precise determination of the
extintion value toward this system. 

It is ineresting to compare the abundances we derived from the edge
fitting with the expected value for the ISM (as reported by
\citealt{wil00}). We find that in the direction of J1808, the Ne/O
abundance is $\sim$0.23, slightly larger than in the ISM ($\sim0.18$),
which maybe be poiting to a Ne-rich environment as observed in other
low mass X-ray binaries (Juett et al. 2001).

\section*{Acknowledgments}

We thank the \XMM\, observatory for promptly assigning us an
observation of \src\, through its Director's Discretionary Time.  We
thank M. Diaz-Trigo, E. Costantini, A. Watts, A. Raassen and
P. Casella for useful suggestions, and J.Poutanen for stimulating
discussions on the pulse formation process in accreting pulsars.  NR
acknowledges support through an NWO Veni Fellowship.

\newcommand{\nat}{Nat}
\newcommand{\mnras}{MNRAS}
\newcommand{\aj}{AJ}
\newcommand{\pasp}{PASP}
\newcommand{\aap}{A\&A}
\newcommand{\apj}{ApJ}
\newcommand{\araa}{ARA\&A}
\newcommand{\apjl}{ApJL}
\newcommand{\apss}{ApSS}
\newcommand{\apjs}{ApJS}
\newcommand{\aaps}{AAPS}
\newcommand{\gca}{GeCoA}

\end{document}